\documentclass{article}
\usepackage[bookmarks=false, hidelinks]{hyperref}
\usepackage{dcase2018,amsmath,graphicx,url,times,booktabs, tabularx}
\usepackage{color}

\title{General-purpose tagging of Freesound audio with AudioSet labels:\\
Task Description, Dataset, and Baseline}

\name{Eduardo Fonseca$^{1}\sthanks{This work is partially supported by the European Union's Horizon 2020 research and innovation programme under grant agreement No 688382 AudioCommons and a Google Faculty Research Award 2017.}$,
      Manoj Plakal$^{2}$,
      Frederic Font$^{1}$, 
      Daniel P. W. Ellis$^{2}$,
      }
\secondlinename{	  
      Xavier Favory$^{1}$, 
      Jordi Pons$^{1}$,
      Xavier Serra$^{1}$
      }
\address{$^1$Music Technology Group, Universitat Pompeu Fabra, Barcelona \{name.surname\}@upf.edu\\          
        $^2$ Google, Inc., New York, NY, USA \{plakal,dpwe\}@google.com \\
 }

\begin{document}

\ninept
\maketitle

\begin{sloppy}

\begin{abstract}
This paper describes Task 2 of the DCASE 2018 Challenge, titled ``General-purpose audio tagging of Freesound content with AudioSet labels''. This task was hosted on the Kaggle platform as ``Freesound General-Purpose Audio Tagging Challenge''. The goal of the task is to build an audio tagging system that can recognize the category of an audio clip from a subset of 41 diverse categories drawn from the AudioSet Ontology. We present the task, the dataset prepared for the competition, and a baseline system.
\end{abstract}


\section{Introduction}
\label{sec:intro}

The sounds in our everyday environment carry a huge amount of information of the events occurring nearby.
Humans are able to recognize and discern many sound events but state-of-the-art automatic processing of sounds by machines still lags far behind. 
Further research is needed to develop robust systems capable of recognizing a wide range of sound events in realistic audio streams~\cite{virtanen2018computational}.
In recent years, the various editions of the DCASE Challenge have provided scenarios for evaluating different computational methods for several sound recognition tasks using common publicly available datasets~\cite{DCASE2017challenge}. 
This paper describes the characteristics, dataset and baseline of DCASE 2018 Task 2 ``General-purpose audio tagging of Freesound content with AudioSet labels''.

There have been two audio tagging tasks in previous DCASE editions, each focused on a specific domain of sounds.
In DCASE 2016 \cite{mesaros2018detection}, the task targeted domestic audio tagging for which the CHiME-Home dataset was used, including 7 sound categories and 6.8h of recordings \cite{foster2015chime}.
In DCASE 2017 \cite{DCASE2017challenge}, the task focused on audio tagging in the context of smart cars, for which a large-scale dataset featuring 17 categories was utilized.
Other popular datasets for sound event classification are the ESC-50 \cite{piczak2015esc} and the UrbanSound8k \cite{salamon2014dataset} datasets. 
The former counts with 50 diverse categories but less than 3h of audio recordings.
The latter is designed to enable urban sound research, featuring 10 categories and almost 9h of audio.
Hence, many of available data resources for sound event classification are domain-specific, and/or of relatively small size.
Recently, however, general-purpose sound event recognizers have gained attention, where a wide range of sounds events are considered, not tied to a specific domain.
This research has been mostly triggered by AudioSet, a large-scale audio dataset structured with an ontology of 632 sound events \cite{gemmeke2017audio}.

In this paper, we focus on general-purpose audio tagging using a dataset of 41 categories and almost 18h of training data.
Specifically, the goal of this task is to build an audio tagging system that can categorize an audio clip as belonging to one of a set of 41 diverse categories drawn from the AudioSet Ontology (related to musical instruments, human sounds, domestic sounds, animals, etc.).
One of the motivations for this task comes from the large amount of user-generated audio content that is available on the web, which can be a resource of great potential for sound recognition related research.
The use of such data for training audio tagging systems poses issues that have not been addressed in previous DCASE Challenges.
In particular, this task deals with user-generated audio clips retrieved from Freesound,\footnote{\url{https://freesound.org}}
which are very diverse in terms of acoustic content, recording techniques, clip duration, etc.
Likewise, these audio clips sometimes feature incomplete and inconsistent user-provided metadata.
To prepare the dataset for this task, some audio clips were manually labeled using the subset of 41 categories, while a larger set of clips was automatically categorized on the basis of their existing user-provided metadata (see Section~\ref{sec:dataset} for more details). 
As a result, the dataset features a small amount of reliable annotations, and a large amount of non-verified annotations that could include a small amount of label noise.

Therefore, this task addresses two main challenges of \textit{i)} recognizing an increased number of diverse sound events, and \textit{ii)} leveraging subsets of training data featuring annotations of varying reliability.
Submissions to this task will provide insight towards the development of broadly-applicable sound event classifiers.
Potential applications include automatic description of multimedia content, and acoustic monitoring applications.
This paper is organized as follows. 
Section \ref{sec:challenge} provides more details about the task and its experimental setup.
Section \ref{sec:dataset} presents in detail the dataset prepared for the task, and Section \ref{sec:baseline} describes a baseline system.
Final remarks are given in Section \ref{sec:conclusion}.

\vspace{-3mm}

\section{Task Setup}
\label{sec:challenge}
\vspace{-1mm}
The goal of this task is to predict the category for each audio clip in a test set.
The task setup is a multiclass classification problem, and hence the systems to be developed in this task can be denoted as \emph{single-tag} audio tagging systems, as illustrated in Fig.~\ref{fig:task_diagram}. 
This task was hosted on Kaggle---a platform for machine learning competitions---and ran from March 30th to July 31st 2018. 
The resources associated to this task (dataset download, submission, and leaderboard) can be found on the Kaggle competition page.\footnote{\url{https://kaggle.com/c/freesound-audio-tagging}\\Note that competition name on Kaggle is abbreviated from the full DCASE task name to ``Freesound General-Purpose Audio Tagging Challenge''.\label{footnote_Kaggle_page}}
\begin{figure}[ht]
  \centering
  \centerline{\includegraphics[width=\columnwidth]{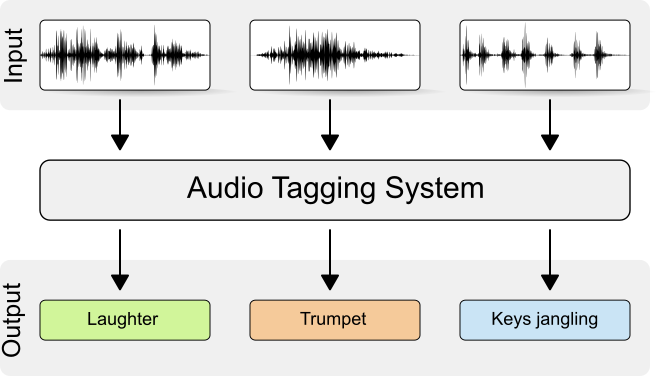}}
  \caption{Overview of a \emph{single-tag} tagging system.}
  \label{fig:task_diagram}
  \vspace{-4mm}
\end{figure}

As described in Section~\ref{sec:dataset}, the audio data for this task are split into a train set and a test set, both made publicly available when the competition launched.
The train set, for which ground-truth annotations were provided, is used for system development while the test set is kept for the evaluation of the resulting systems.
The test set, whose true labels were not released, is further divided into two divisions: \textit{i)} 19\% of the test samples are used to calculate the \textit{public} leaderboard (providing a live ranking of all participants), and \textit{ii)} the remaining 81\% feeds the \textit{private} leaderboard, used for the final ranking which is revealed only when the competition ends.

\subsection{Evaluation Metric and Competition Rules}

The task used mean Average Precision @ 3 (mAP@3) as the evaluation metric, as defined in the Evaluation section of the competition page.\textsuperscript{\ref{footnote_Kaggle_page}}
%
This metric accepts up to three ranked predicted labels for each audio clip in the test set, and gives full credit if the correct label occurs first, with lesser credit for correct label predictions in second or third place.

Participants were required to run their systems on the test set and submit the system output---the predicted labels---in a comma-separated data file (CSV). 
Participants could submit a maximum of two submissions per day, and select two \emph{final} submissions to be considered for the ranking. 
Additionally, participants were encouraged to submit a technical report describing their systems.
A detailed description of the task rules can be found in the Rules section of the competition page,\textsuperscript{\ref{footnote_Kaggle_page}} and the most important points are summarized in the DCASE Challenge page.\footnote{\url{http://dcase.community/challenge2018/task-general-purpose-audio-tagging\#task-rules}} 

\vspace{-2mm}
\subsection{Judges' Award}
To complement the leaderboard results of the mAP-based ranking, the organizers of the task introduced a complementary Judges' Award to promote submissions using novel, problem-specific and efficient approaches. Details about the Judges' Award rules and requirements can be found in the Discussion section of the competition page.\footnote{\url{https://www.kaggle.com/c/freesound-audio-tagging/discussion/59932}}
\vspace{-3mm}

\section{Dataset}
\label{sec:dataset}

The dataset used for the task was prepared by the task organizers during the months previous to the start of the competition, and is called ``Freesound Dataset Kaggle 2018'' (or \textit{FSDKaggle2018} for short). 
FSDKaggle2018 is in fact a reduced subset of \textit{FSD}, which is a large-scale, general-purpose open audio dataset that is currently under development. 
More details about FSD can be found in~\cite{Fonseca2017freesound}. 
The following subsections describe the creation process of FSDKaggle2018.

\vspace{-2mm}
\subsection{Source of Audio Content}
\label{ssec:source}
FSDKaggle2018 is composed of audio content collected from Freesound.
Freesound is a sound sharing site developed and maintained by the Music Technology Group~\cite{font2013freesound}. 
At the time of writing, Freesound hosts more than 380,000 sounds uploaded by its community of users.
Freesound content is very heterogeneous, including sounds from a wide range of real-world environments, from musical and human-generated sounds to animal sounds or artificially generated sound effects.
The authors of the sounds uploaded to Freesound are asked to provide some basic metadata (e.g., tags, title and textual descriptions). This metadata is then used for searching and browsing and is also very valuable for research purposes.
All the content in Freesound is released under Creative Commons licenses which facilitate its sharing and reuse.
Since sounds are uploaded by thousands of users across the globe, recording scenarios and techniques can vary widely. We hypothesize that this fact makes Freesound content representative of real-world situations.

\vspace{-2mm}
\subsection{Annotation Procedure}
\label{ssec:annotation}
FSDKaggle2018 is organized using categories of the AudioSet Ontology.
As a first step, we did a mapping of 268,261 Freesound clips to their corresponding AudioSet categories. To do that, we assigned a number of Freesound tags to almost all of the 632 AudioSet categories and, for each category, we selected audio clips from Freesound tagged with at least one of these tags.
This process led to a number of automatically generated \textit{candidate annotations} that express the potential presence of a sound category in an audio clip. 
These annotations are at the clip level and hence can be considered weak labels. 
However, some audio files are specific sound examples of the category under consideration, where the acoustic signal fills almost the entirety of the file, which could arguably be considered as strong labels.
 
In order to validate the candidate annotations, we used \emph{Freesound Datasets},\footnote{\url{https://datasets.freesound.org}} an online platform for the collaborative creation of open audio datasets developed at the Music Technology Group.
We deployed a validation task in which Freesound Datasets users can manually verify the presence or absence of a candidate sound category in an audio clip with a \emph{rating} mechanism.
For every sound category, users first go through a training phase to get familiar with the category, read its description provided by AudioSet, and listen to some selected sound examples. 
Then, users are presented with a series of audio clips, and prompted the question: \textit{Is $<$category$>$ present in the following sounds?} Users must select one of the response types listed in Table~\ref{tab:responsetypes}. Along with the audio clips, users are also given links to the corresponding Freesound sound pages where the original tags and descriptions are available and can be used as an aid for the validation process.
Participants in the validation task included voluntaries from the Freesound community as well as researchers and students from the Music Technology Group.

\begin{table*}[p]
\centering
\caption{Response types for the validation task. Users must select one to answer the question: \textit{Is $<$category$>$ present in the following sounds?}}
\vspace{0.2cm}
\begin{tabular}{ll}

\textbf{Response type}	    &  \textbf{Meaning}   \\ \hline 
Present and 			&The type of sound described is \textbf{clearly present} and \textbf{predominant}.\\
predominant	(PP)			&This means there are no other types of sound, with the exception of low/mild background noise.\\	
\hline 
Present but not 		&The type of sound described is \textbf{present}, but the audio clip also\\
predominant (PNP)			& \textbf{contains other salient types of sound and/or strong background noise}.\\
\hline 
Not Present	(NP)			&The type of sound described is \textbf{not present} in the audio clip. \\
\hline 
Unsure (U)					&\textbf{I am not sure} whether the type of sound described is present or not.\\

\end{tabular}
\label{tab:responsetypes}
\end{table*}

Among the various features implemented in the validation task, it is worth mentioning the utilization of quality control mechanisms such as the periodic inclusion of verification clips to test the reliability of the submitted responses.
Likewise, in order to choose which audio clips to present to each user, we adopt a prioritization scheme that considers inter-annotator agreement. 
More specifically, each candidate annotation is presented to several users (i.e., annotators) until agreement is attained by two different users on a response type. 
When a candidate annotation reaches an agreement status, it is considered validated and is no longer presented to other users.

\vspace{-2mm}
\subsection{Dataset Curation}
\label{ssec:curation}

After generating candidate annotations and collecting user ratings in the Freesound Datasets platform, each candidate annotation had a particular distribution of ratings \{PP, PNP, NP, U\} (see Table~\ref{tab:responsetypes}). Then, a curation step was carried out to select which categories and audio clips to be finally included in FSDKaggle2018. Considering all annotations, two annotation subsets were created for each sound category:


\begin{itemize}
\item \textbf{Manually-verified annotations}: composed of those annotations rated only as PP (a great majority with inter-annotator agreement but not all of them, hence PP \& PP or single PP).

\item \textbf{Non-verified annotations}: composed mainly of the \emph{un-rated} candidate annotations, and complemented with a small amount of rated annotations. 
This small amount of rated annotations can include any rating distribution except \textit{i)} those corresponding to the manually-verified portion, and \textit{ii)} those that clearly denote an incorrect mapping (e.g., NP, NP \& U, etc.).

\end{itemize}

For each sound category, a quality estimate \emph{QE} for the non-verified annotations can be computed according to (\ref{eqn:QE})
\begin{equation}
  \label{eqn:QE}
    QE = \frac{\#PP + \#PNP}{\#PP + \#PNP + \#NP + \#U} 
\end{equation}
where $\#X$ denotes the number of ratings of type $X$ gathered in the category. 

Next, a number of restrictions were applied sequentially to the categories and/or the audio clips within them.
First, we discarded all categories not belonging to leaf nodes of the AudioSet hierarchy, leaving a total of 474 categories.
Then, we removed audio clips shorter than 300ms and longer than 30s, as well as those clips with Creative Commons \textit{Non-commercial} or \textit{Sampling+} licenses.
All sound categories that, after the previous filtering, did not have \textit{i)} a minimum of number of manually-verified annotations, and \textit{ii)} a minimum number of manually-verified + non-verified annotations, were discarded.
Note that in order to accept the non-verified annotations in a category, a minimum QE was required (see Section~\ref{ssec:description}). 

We observed that quite a few leaf sound categories were discarded because they did not have sufficient number of clips.
In some of these cases, making use of the hierarchical relationships in AudioSet, we decided to aggregate the content of these leaf categories together with that of their immediate parents in order to create new candidate parent categories.
Similar requirements (in terms of QE and amount of data) were applied to these newly formed categories for them to be accepted in the raw version of FSDKaggle2018.

After this process, an analysis was carried out in terms of \textit{i)} number of \textit{in-domain}\footnote{Considering only the set of valid categories at this point of the process, instead of all the AudioSet categories.} candidate annotations per audio clip and \textit{ii)} semantic aspect of the resulting categories.
The analysis revealed that the vast majority of the audio clips presented a single candidate annotation and, for the sake of simplicity, we decided to discard audio clips with multiple annotations.\footnote{Note that the automatically generated candidate annotations depend on the user-generated tags in Freesound and on the mapping to the AudioSet Ontology. Hence their reliability relies on the subsequent validation process.}
We also discarded a few categories with somewhat abstract or vague meaning like ``Recording'' or ``Effect unit''.

Finally, the audio clips with manually-verified annotations for every category were split into roughly 70\%/30\% for train and test sets. The split was carried out considering, whenever possible, clip origin (by using part of the Freesound metadata) and clip duration (so as to have short and long clips in both sets).
Then, we complemented the manually-verified portion of the train set with the non-verified annotations. 
This addition was performed such that the maximum number of clips per category was 300 in order to mitigate data imbalance among categories.
The dataset curation resulted in the selected 11,073 sounds/annotations organized with 41 AudioSet categories.



\vspace{-2mm}
\subsection{Dataset Description}
\label{ssec:description}
FSDKaggle2018 contains a total of 11,073 files provided as uncompressed PCM 16 bit, 44.1 kHz, mono audio files.
All audio clips are released under either Creative Commons \emph{Attribution} or \emph{Zero} licenses.
The clips are unequally distributed in the 41 categories of the AudioSet Ontology listed in Table \ref{tab:categories}.
The dataset most relevant characteristics are as follows:
\begin{itemize}

\item Audio clips are annotated with a single ground truth label.

\item The duration of the audio clips ranges from 300ms to 30s due to the diversity of the sound categories and the preferences of Freesound users when recording sounds. 

\item The dataset is split into a train set and a test set. 

\item The \textbf{train set} is meant to be for system development and includes 9473 audio clips unequally distributed among 41 categories. The minimum number of audio clips per category in the train set is 94, and the maximum is 300. The total duration of the train set is almost 18h.

\item Out of the 9473 clips from the train set, 3710 have manually-verified annotations and 5763 have non-verified annotations. The latter are properly flagged so that participants can opt to use this information during the development of their systems.

\item The \textbf{test set} is composed of 1600 clips with manually-verified annotations and with a similar category distribution to that of the manually-verified portion of the train set.
The minimum number of manually-verified audio clips per category in the test set is 25, and the maximum is 110.
The test set is complemented with 7800 \textit{padding} clips.\footnote{Hence, the dataset available from Kaggle contains 18,873 audio files.}
These clips, which are not used for scoring the systems, are added to prevent undesired practices (considering that the test set was made publicly available when the competition launched).


\end{itemize}

\begin{table*}[p]
\centering
\caption{Categories composing FSDKaggle2018, along with the number of clips and time (in minutes, rounded) in the train set. Per-category AP@3 achieved by the baseline system is reported using all the test files for every category (i.e., not following the public/private splits of the Kaggle leaderboard).}
\vspace{0.2cm}
\begin{tabular}{@{}c@{} c c c |@{}c@{} c c c |@{}c@{} c c c}

\textbf{Name}    &\textbf{clips}   &\textbf{time}  &\textbf{AP@3}  &\textbf{Name}   &\textbf{clips}   &\textbf{time}  &\textbf{AP@3}  &\textbf{Name}   &\textbf{clips}   &\textbf{time}  &\textbf{AP@3} \\ 
\hline
Acoustic guitar 	&300  				&52				&0.67		&Electric piano			&150  				&25				&0.75      &Microwave oven 		&146  				&25				&0.56 \\
Applause		 	&300  				&58				&0.98		&Fart 					&300  				&18				&0.65      &Oboe					&299  				&15				&0.88 \\
Bark			 	&239  				&45				&0.85		&Finger snapping		&117  				&6				&0.71      &Saxophone 			&300  				&34				&0.84 \\
Bass drum 			&300  				&13				&0.55		&Fireworks 				&300  				&48				&0.61      &Scissors				&95  				&16				&0.37 \\
Burping,eructation 	&210  				&12				&0.71		&Flute 					&300  				&46				&0.90      &Shatter 				&300  				&26				&0.70 \\
Bus 				&109  				&28				&0.53		&Glockenspiel 			&94  				&8				&0.59      &Snare drum 			&300  				&18				&0.30 \\
Cello 				&300  				&37				&0.86		&Gong 					&292  				&42				&0.81      &Squeak 				&300  				&38				&0.16 \\
Chime 				&115  				&24				&0.79		&Gunshot,gunfire 		&147  				&11				&0.16      &Tambourine 			&221  				&10				&0.78 \\
Clarinet 			&300  				&35				&0.96		&Harmonica 				&165  				&19				&0.86      &Tearing 				&300  				&39				&0.94 \\
Computer keyboard 	&119  				&23				&0.54		&Hi-hat 				&300  				&19				&0.53      &Telephone 			&120  				&16				&0.65 \\
Cough 				&243  				&22				&0.69		&Keys jangling 			&139  				&19				&0.76      &Trumpet 				&300  				&28				&0.84 \\
Cowbell 			&191  				&11				&0.58		&Knock 					&279  				&19				&0.89      &Violin,fiddle		&300  				&27				&0.73 \\
Double bass 		&300  				&17				&0.69		&Laughter				&300  				&36				&0.96      &Writing 				&270  				&48				&0.66 \\
Drawer open,close 	&158  				&18				&0.05		&Meow 					&155  				&19				&0.82       & & & & \\

\end{tabular}
\label{tab:categories}
\end{table*}

As mentioned in Section~\ref{ssec:curation}, all \textbf{manually-verified annotations} are annotations validated as PP (Present and Predominant).
This means that, in most cases, there is no additional acoustic material other than the labeled category.
In few cases, there may be some additional sound events, but these additional events will be \emph{out-of-domain}, i.e., they won't belong to any of the 41 AudioSet categories of FSDKaggle2018. 
The \textbf{non-verified annotations} have a QE of \textit{at least} 65\% in each category. 
This means that some of them are most probably inaccurate.
It can happen that audio clips corresponding to some of the non-verified annotations present several sound sources (even though only one label is provided as ground truth). These additional sources are typically out-of-domain, but in a few cases they could be within the domain. 
Fig. \ref{fig:annotations} shows the distribution of manually-verified and non-verified annotations per category in the train set.
\vspace{-4mm}


\begin{figure*}[p]
  \centering
  \centerline{\includegraphics[width=0.95\textwidth]{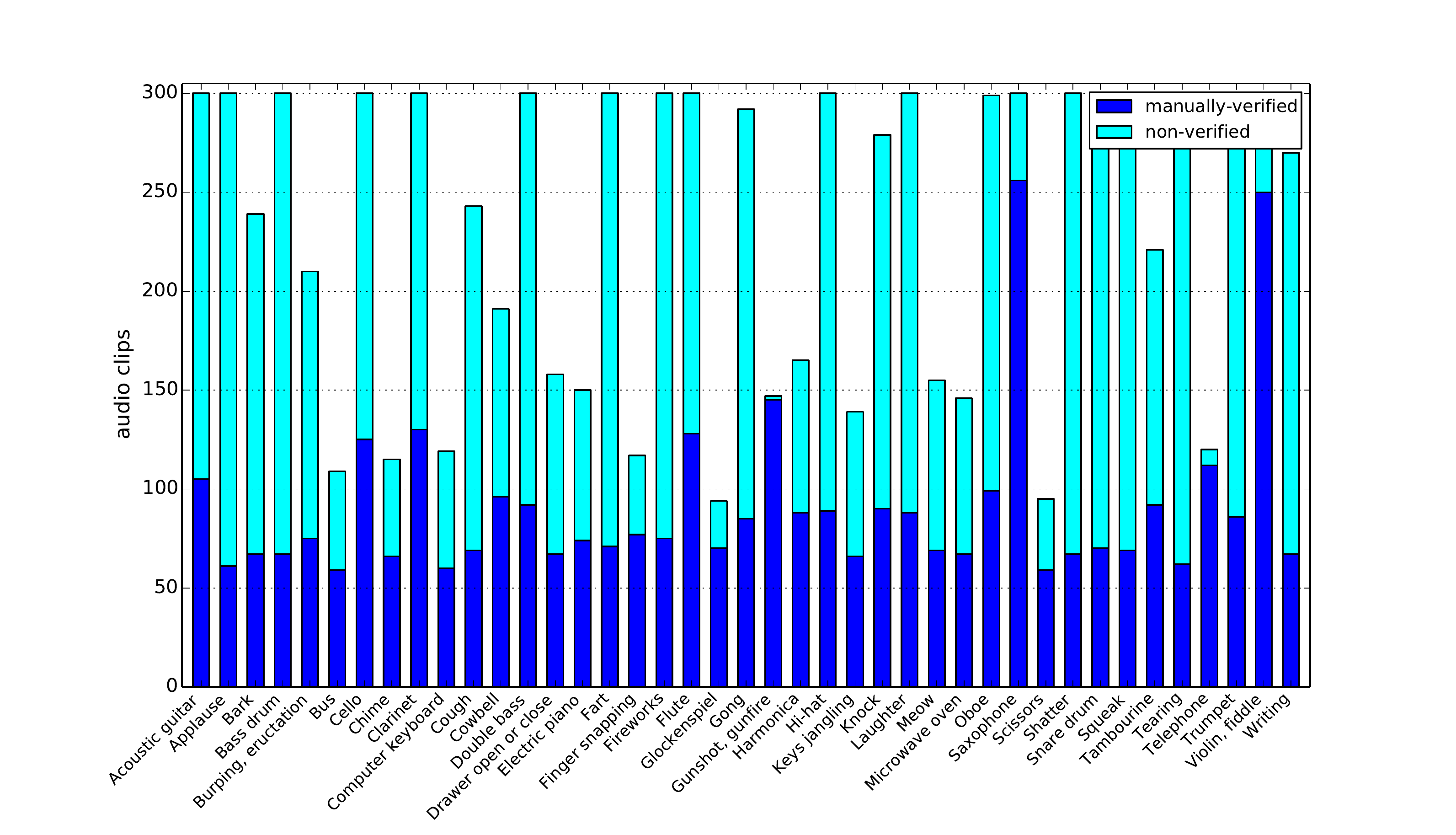}}
  \caption{Distribution of manually-verified and non-verified annotations per category in the train set.}
  \label{fig:annotations}
\end{figure*}

\section{Baseline System}
\label{sec:baseline}
In recent years, convolutional and recurrent neural networks (CNNs, RNNs, CRNNs) have achieved state-of-the-art performance for audio tagging and event detection in DCASE Challenges \cite{dcase2016task4results,dcase2017task4results}, and deep CNNs have been shown to work well with very large datasets \cite{hershey2017cnn,aytar2016soundnet}, outpacing simpler models.

Our baseline uses a relatively shallow 3-layer CNN with log-mel spectrogram input features and a 41-way softmax classifier layer, described in detail in the public release \footnote{\url{https://github.com/DCASE-REPO/dcase2018_baseline/tree/master/task2}}. Incoming audio (always 44.1 kHz mono) is divided into overlapping windows of size 0.25s with a hop of 0.125s. These windows are decomposed with a short-time Fourier transform using 25ms windows every 10ms. The resulting spectrogram is mapped into 64 mel-spaced frequency bins, and the magnitude of each bin is log-transformed after adding a small offset to avoid numerical issues. The model consists of three 2-D convolutional layers and alternating max-pooling layers, followed by a softmax classifier layer. Predictions are obtained for a clip of arbitrary length by running the model over 0.25s-wide windows every 0.125s, and averaging all the window-level predictions to obtain a clip-level prediction.

The baseline achieves an mAP@3 of 0.70 on the entire test set (0.70 and 0.69 on the public and private Kaggle leaderboard splits, respectively) after training for 5 epochs on the train set. Per-category AP@3 is reported in Table~\ref{tab:categories}.

\vspace{-2mm}

\section{Conclusion}
\label{sec:conclusion}
In this paper we have described the task setup, dataset, and baseline of DCASE 2018 Task 2 ``General-purpose audio tagging of Freesound content with AudioSet labels''. 
This task was hosted on the Kaggle platform as ``Freesound General-Purpose Audio Tagging Challenge'' and ran from March 30th to July 31st 2018. 
The main focus of the paper is the description of FSDKaggle2018, the dataset we prepared for the task. 
FSDKaggle2018 presents the particularities of having subsets of training data with annotations of varying reliability as well as featuring variable-length audio clips, both novel challenges in DCASE competitions. The dataset is currently available on the Kaggle competition page, and future updates of the dataset (including ground-truth data for the test set and extra associated Freesound metadata) will be made publicly available in the Freesound Datasets platform.
Through FSDKaggle2018 and the provided baseline system, this competition intends to foster open research in sound event recognition.

\vspace{-2mm}

\section{ACKNOWLEDGMENT}
\label{sec:ack}
We thank Addison Howard and Walter Reade of Kaggle for their invaluable assistance with the task design and Kaggle platform, and everyone who contributed to FSDKaggle2018 with annotations. 
Eduardo Fonseca is also grateful for the GPU donated by NVidia.




\bibliographystyle{IEEEtran}
\bibliography{refs}
%
%
%
%
%
%
%
%
%

\end{sloppy}
\end{document}